# Information literacy development and assessment at school level: a systematic review of the literature

Luz Chourio-Acevedo[a,*], Jacqueline Köhler[a], Carla Coscarelli[b], Daniel Gacitúa[a], Verónica Proaño-Ríos[a,c], Roberto González-Ibáñez[a]

[a] Departamento de Ingeniería Informática, Universidad de Santiago de Chile, Avenida Ecuador 3659, 9160000, Estación Central, Chile

[b] Faculdade de Letras, Universidade Federal de Minas Gerais, Belo Horizonte, Brazil

[c] Departamento de Ciencias Exactas, Universidad de las Fuerzas Armadas—ESPE, Sangolquí, Ecuador

[*] Corresponding author. Email addresses: luz.chourio@usach.cl

**Authors note**

The work described in this article was partially supported by the TUTELAGE project, funded by the National Agency for Research and Development (ANID) FONDECYT Regular [grant no. 1201610]; ANID Scholarship Program Doctorado Nacional [grant no. 2019-21192162]; ANID Doctorado Becas Chile [grant no. 7608-2020]; Coordenação de Aperfeiçoamento de Pessoal de Nível Superior–Brasil (CAPES)–Finance Code 001 Processo: CAPES-PRINT-88887.373484/2019-00, the Vicerrectoría de Postgrado USACH (Beca de Excelencia para Extranjeros 2019) and Secretaría Nacional de Educación Superior, Ciencia,

**Declaration of Interest**

The authors declare that they have no known competing financial interests or personal relationships that could have appeared to influence the work reported in this paper

**Author contributions**

Luz Chourio-Acevedo: Conceptualization, Methodology, Formal analysis, Investigation, Data curation, Roles/Writing - original draft, Writing – review & editing, Visualization, Supervision, Project administration, Funding acquisition. Jacqueline Köhler: Conceptualization, Methodology, Formal analysis, Investigation, Resources, Data curation, Roles/Writing - original draft; Writing - review & editing; Visualization; Project administration; Funding acquisition. Carla Coscarelli: Methodology; Formal analysis; Investigation; Resources; Data curation; Roles/Writing - original draft; Writing - review & editing; Visualization; Supervision; Project administration; Funding acquisition. Daniel Gacitúa: Methodology; Software; Formal analysis; Investigation; Data curation; Roles/Writing - original draft; Writing - review & editing; Visualization; Funding acquisition. Verónica Proaño-Ríos: Methodology; Formal analysis; Investigation; Data curation; Roles/Writing – original draft; Writing - review & editing; Visualization; Funding acquisition. Roberto González-Ibáñez: Conceptualization; Validation; Roles/Writing – original draft; Writing - review & editing; Funding acquisition.

**Abstract**

Information literacy (IL) involves a group of competences and fundamental skills in the 21st century. Today, society operates around information, which is challenging considering the vast amount of content available online. People must be capable of searching, critically assessing, making sense of, and communicating information. This set of competences must be properly developed since childhood, especially if considering early age access to online resources. To better understand the evolution and current status of IL development and assessment at school (K-12) level, we conducted a systematic literature review based on the guidelines established by the PRISMA statement. Our review led us to an initial set of 1,234 articles, from which 53 passed the inclusion criteria. These articles were used to address six researchquestions focused on IL definitions, skills, standards, and assessment tools.Our review shows IL evolution over the years and how it has been formalisedthrough definitions and standards. These findings reveal key gaps that mustbe addressed in order to advance the field further.

**Information literacy development and assessment at school level:**

**a systematic review of the literature**

The world has experienced great changes since the arrival of the World Wide Web and massively available Information and Communication Technologies (ICTs), which is evidenced in how processes and functions of the information age are concentrated around networks (Castells, 2010). Education is not an exception, and lots of research has been done since Delors claimed that skills necessary to use new technologies should be included into the curricula (Delors, 1998). As a consequence, organisations, such as UNESCO, have considered this subject (Horton, 2008), and OECD has included ICT skills in PISA and PIAAC evaluations (Kankaraš et al., 2016; He et al., 2017; OECD & Statistics Canada, 2000). In this context, Information Literacy (IL) is of special interest, since "information literate people know how to find, evaluate, and use information effectively to solve a particular problem or make a decision" (American Library Association, 1989, para. 19). It is crucial that people adapt to changes in today's world and that they make effective use of the information available (Delors, 1998; Horton, 2008; American Library Association, 1989).

Given the importance of IL in today's world, Computer Science must contribute to this field with tools to support the development of the associated skills on users. As IL is a rather new research area, it is not easy to find a guide that compiles the generated knowledge. Therefore, this work aims to contribute to this task by the means of a systematic literature review focused on the evolution and current status of IL teaching and assessment at school (K-12) level. Works on this subject come mainly from Library and Information Science, and fields, such as Computer Science, have contributed by adding new perspectives and creating new tools.

A systematic literature review "attempts to collate all empirical evidence that fits pre-specified eligibility criteria in order to answer a specific research question. It uses explicit, systematic methods that are selected with a view to minimising bias, thus providing more reliable findings from which conclusions can be drawn and decisions made" (Higgins et al., 2011, p. 27).

In this work, we aim to answer the following question: How have information literacy development and assessment been addressed at school level? This is guided by 6 research questions (RQ):

1. How has information literacy been defined over time?
2. What are the skills or competences related to IL?
3. What tools and techniques are proposed in the literature for assessing information skills of school students?
4. What technological tools are discussed in the literature to develop information skills on school students?
5. How does the use of technological tools impact on the development of information skills?
6. Which IL standards for k-12 students are addressed in the literature?

**Method**

This systematic literature review has been conducted according to the guidelines proposed by the Cochrane Handbook for Systematic Reviews (Higgins et al., 2011) and the PRISMA flow diagram (Page et al., 2021) so that the research can be reproduced and validated, and to ensure that this work has all the necessary items.

The following subsections describe the results of each of the steps performed in this literature review.

**Information Sources**

Information literacy has been studied by different fields related to Information Science around the world. Although English can be considered a standard language in terms of academic literature, Spanish is the second most common language in academia (Badillo, 2021). Therefore, we inquired about the Scopus and Web of Science (WoS) academic literature collections, which list most of the main worldwide journals and proceedings; the Scielo literature collection, which lists most relevant Latin-American research; and specialised journals and conferences, namely JIL (Journal of Information Literacy), ECIL (European Conference on Information Literacy) and ICIL (International Conference on Information Literacy).

Two search queries were considered during the search process:

1. ("information literacy" OR "online inquiry" OR "online search" OR "information search" OR "information fluency") AND (competences OR skills) AND ((student* OR teacher) AND ((primary OR elementary OR secondary OR high) AND (education OR school)) OR k-12)

2. ("information literacy" OR "online inquiry" OR "online search" OR "information search" OR "information fluency") AND (competences OR skills) AND ((training OR evaluation OR assessment OR standard) AND ((primary OR secondary OR elementary OR high) OR k-12))

The equivalent search queries for Ibero-American sources were:

1. ("alfabetización informacional" OR "alfabetización en información" OR "desarrollo de (habilidades informativas" OR "DHI" OR "competencias informacionales") AND (estudiante* OR alumno* OR profesor* o docente*) AND (educación OR escuela OR colegio OR liceo) AND (primaria OR secundaria OR básica OR media OR escolar)

2. ("alfabetización informacional" OR "alfabetización en información" OR "desarrollo de (habilidades informativas" OR "DHI" OR "competencias informacionales") AND (habilidades OR competencias) AND (enseñanza OR evaluación OR estándar) AND (educación OR escuela OR colegio OR liceo) AND (primaria OR secundaria OR básica OR media OR escolar)

**Inclusion and Exclusion Criteria**

The following inclusion criteria (IC) were considered for filtering retrieved articles, first applied to document titles and then, to abstracts:

1. Documents containing concepts, experiences or IL models oriented toward school students.
2. Documents describing studies regarding IL in school students.
3. Documents regarding IL training or skills of elementary or secondary school teachers.

Similarly, the following exclusion criteria (EC) were followed:

1. Multimedia documents.
2. IL related documents focused on a population different from school students.
3. Documents written in languages other than English or Spanish.
4. Duplicated documents.

It must be stated that no time limits were established, due to the nature of RQ1.

**Quality Assessment**

The selection process was followed by a quality assessment of the remaining articles, which is based on the criteria proposed by Kmet, Cook, and Lee (2004):

1. Question / objective sufficiently described?

        2.        Study design evident and appropriate?

        3.        Is the method of selecting the source of information described and appropriate?

        4.        Is the sample size appropriate?

        5.        Is the sampling strategy described and justified?

        6.        Are data collection methods clearly described and systematic?

        7.        Is data analysis clearly described and systematic?

        8.        Are the conclusions supported by the results?

**Data Selection and Extraction**

The data selection and extraction processes followed the guidelines established by PRISMA (Page et al., 2021), which states the stages: Identification, Screening, and Included. Figure 1 details this process, illustrating how the article filtering and selection was conducted, whereas Table 1 summarises the results of each of the stages.

The Identification stage was conducted using the Parsifal tool, where both the search strings and the retrieved articles from each source were stored. A total of 2,378 articles were retrieved, out of which 1,234 remained after the duplicate identification.

The Screening stage involved filters by title and abstract, considering the inclusion and exclusion criteria. These filters were applied by two peers, who discussed the differences to achieve an agreement. It must be noted that special care was taken in order to avoid that an article was assessed by the same couple of reviewers in the different stages. For the title and abstract filters, each reviewer assessed each article as Accepted or Rejected. 440 articles remained after filtering by title, and this number was reduced to 252 after filtering by abstract. However, only 209 of the remaining articles could be downloaded as a consequence of access restrictions for the researchers' institutions libraries. Afterwards, the previously described quality assessment was employed. Each article was evaluated by two peers, each of whom assigned a score of 1 if the criterion was accomplished, and 0 otherwise. Therefore, each article obtained a total score ranging from 0 to 8. If the score given by the reviewers differed in 3 or more points, they discussed the differences in order to achieve an agreement. The final score for each work was calculated as the average score. Also, Cohen's Kappa was used to assess the inter-rater reliability for each of the quality assessment criteria, as shown in Table 2. It can be observed that substantial agreement was achieved for all of the criteria (Landis & Koch, 1977).

Finally, during the Included stage, the articles that best fulfilled the criteria were selected. These correspond to articles in the highest quality score quartile (≥7.0) (Figure 3), which are listed in Table 3.

Selected works cover a time-span from 2001 to 2020 (although articles written as early as 1984 were retrieved). The number of articles clearly tends to increase over the years, with a notorious peak in 2019. It must be noted that this study only considers articles published up to June 2020. Most of the selected studies concentrate in Europe, specially in Germany, Norway and Spain. The second place is held by Asia, America holds the third place, followed by Oceania and Africa. The data show that there is little research at the elementary school level (18.87%), with the majority of the research concentrated on high school (67.92%). Most selected studies are quantitative (43.64%), due to the fact that it is more feasible to apply and scale up these kinds of studies on K-12 students. Nevertheless, there is an important amount of mixed studies (32.73%), showing that qualitative research is also an important tool for studying Information Literacy related skills. Although the main study subjects are students, some studies involved teachers, librarians and parents, since they have a great influence on students' education.

## Results

This section presents the analysis based on the 6 proposed research questions.

**RQ1. How has information literacy been defined over time?**

A definition of Information Literacy is frequently provided by the authors (13 articles), although in the majority of the articles (n=18) concepts formulated by other authors and from standards are quoted.

The concepts bring features that are recurrent in the definitions and that can be taken as key pieces for the conceptualisation of Information Literacy. Basically, Information Literacy (IL) has been described as "the ability to locate, access, search, evaluate and use information in different contexts" (Chang, Foo & Majid 2014, p.1). It is important to highlight that the various contexts in which IL is involved include physical as well as digital environments (Claro et al., 2012).

The idea of effective use of information is mentioned in many articles, as we can see in Bebbington and Vellino (2015) and Kovalik, Yutzey, and Piazza (2013):

- "Information literacy is the ability to know when there is a need for information, to be able to identify, locate, evaluate, and then use that information effectively to make informed decisions" (Bebbington & Vellino, 2015, p.8).
- "Information literacy is an important area of research because of the abundance of readily available information that needs to be located, evaluated, and analysed to be used effectively and appropriately" (Kovalik, Yutzey & Piazza 2013, p.2).

It is important to go beyond IL as "collecting information and rearranging facts" (Gordon (2002), quoted in Kovalik, Yutzey & Piazza (2013, p.2)). In Kovalik, Yutzey and Piazza (2013), attention is targeted to the necessity of helping students to reach and engage in higher-order thinking, which leads to another frequently mentioned aspect, the critical analysis of the information (Herring, 2006):

- "to critically analyse this information" (List, Brante & Klee, 2020, p.2).
- "ability to be creative and innovative, solving problems and thinking critically with the Internet and computers" (Aesaert et al. (2015), quoted in Zhang and Zhu (2016, 580)).
- "collaborate, create and share contents and create knowledge; in an effective, efficient, proper, critical, creative, autonomous, flexible, ethical and reflexive way[1]" (Martínez-Piñeiro, Gewerc, & Rodríguez-Groba, 2019, p.4).
- "and uses information in critical thinking and problem solving" (Isfandyari-Moghaddam & Kashi-Nahanji, 2011, p.619).
- become "critical users of information and creative producers of knowledge" (Bowler, Large & Rejskind 2001, p.205).

Previous definitions are augmented by aspects such as problem solving (Martínez-Piñeiro, Gewerc, & Rodríguez-Groba, 2019), communication (List, Brante & Klee, 2020; Zhang & Zhu, 2016; Claro et al., 2012), information usage (Herring 2006; List, Brante & Klee 2020; Zhang & Zhu, 2016; Claro et al., 2012; Chang, Foo & Majid 2014; Bebbington & Vellino, 2015; Isfandyari-Moghaddam & Kashi-Nahanji, 2011; Chen, Chen, & Ma, 2017; Heinström, 2006; Kovalik, Yutzey & Piazza, 2013; Álvarez & Mercè, 2015; Smith et al., 2013; Yu et al., 2011; Ngo, Pickard & Walton, 2019; Azura-Mokhtar, Majid & Foo, 2007; Stanoevska-Slabeva et al., 2017; Aillerie, 2019; Chu, Tse & Chow, 2011), information creation (List, Brante & Klee, 2020; Martínez-Piñeiro, Gewerc & Rodríguez-Groba, 2019), interaction (Zhang & Zhu, 2016; Ngo, Pickard & Walton, 2019) and learning (Herring, 2006; List, Brante & Klee, 2020; Zhang & Zhu, 2016; Chen & Chen, 2016; Bielba-Calvo et al., 2015; Chen, Chen & Ma, 2017; Ngo, Pickard & Walton, 2019).

For several authors (Herring, 2006; List, Brante & Klee, 2020; Zhang & Zhu, 2016; Chen & Chen, 2016; Bielba-Calvo et al., 2015; Chen, Chen & Ma, 2017; Ngo, Pickard &

Walton, 2019) Information Literacy is related to an opportunity to learn, as we can see in Chen and Chen (2016, p.508): "information literacy is using information creatively and reflectively, so that learners' learning experiences can be expanded". The user will not only search for and access information, but will also transform it into knowledge (as Bielba-Calvo et al. (2015, p.126) quote from Ministerio de Educación, Cultura y Deporte (2013, p.199)). Thus, ICT can be undestood as an "essential element to inform, learn and communicate[2]" (Bielba-Calvo et al., 2015, p.126), taken from Ministerio de Educación, Cultura y Deporte (2013, p.21).

IL is important to many domains of human action, such as "work, leisure, participation, learning, socialising, consuming, and empowerment" (as Martínez-Piñeiro, Gewerc & Rodríguez-Groba (2019, p.4), quoted from Ferrari (2012, p.30)). Since this is an interdisciplinary subject, a similar set of skills is gathered in different areas, such as:

- ICT literacy (Claro et al., 2012), digital literacy (List, Brante & Klee, 2020) and media literacy (Zhang & Zhu, 2016).
- "The definitions of ICT literacy, ICT competence, digital literacy and digital competence are also closely related to digital media literacy. According to ETS (2007), ICT literacy is using digital technology, communications tools and/or networks to access, manage, integrate, evaluate and create information in order to function in a knowledge society" (Zhang & Zhu, 2016, p.580).

According to Twing (2013), the concept of digital literacy is similar to that of IL: "digital literacy is the ability to operate effectively as a citizen in the twenty first Century, including having an understanding of the nature of digital technology and the impact of digital identities, being able to interact safely in the digital world, being able to locate, organise, understand, evaluate, analyse and (re)present information using digital technology" (Twing (2013), quoted in Zhang & Zhu (2016, p.580)). These ideas include the abilities mentioned in other definitions such as "locate, organise, understand, evaluate, analyse and (re)present information" (Zhang & Zhu, 2016, p.580).

Although not included in this literature review definitions of IL can be tracked down as early as the 1970s. By then, (Zurkowski, 1974), referred to information literates to those that "have learned techniques and skills for utilising a wide range of information tools as well as primary sources in moulding information solutions to their problems" (p. 6). The concept of IL precedes the digital age and the Web, yet over time some authors treat IL and other types of modern literacies alike.

Information literacy is related to all the communication situations in life. It is definitely not a passive situation but involves communicative circumstances in which the user is requested to be participating and proactive. So, in order to be prepared for personal and professional life, students need to be information literate.

**RQ2. What are the skills or competences related to IL?**

In order to make this analysis, verbs used by the authors when describing the abilities that students need to develop to be considered literate were taken into account. After obtaining the list of skills with their respective frequencies, those skills with 5 or more occurrences were selected, resulting in a list of 22 verbs (Figure 3) with a total of 322 occurrences when considered altogether. Those verbs were then grouped by similarity and organised into four categories, consistently with other studies such as (Leu et al., 2014):

- **Locate** relates to the ones that bring the concept of accessing information.
- **Understand** reunite the verbs related to processing the information. Verbs as comprehend, infer, memorise and conceptualise, although expected were not found in the articles we analysed, when skills were mentioned.

- **Use** assembles the verbs that describe how the information was used or with what purpose.
- **Evaluate** joins actions related to the information's analysis, selection, and evaluation.

Figure 3 shows these 4 skill categories along with their total frequencies and can be interpreted as follows: the most mentioned skills are those associated with the Identification category. The sum of the frequencies of these skills is 86, as detailed below the category name. The Identification category is followed by the skills related to Evaluation, with a total frequency of 83. The next group of skills is associated with Information Usage, with a total frequency of 72 and, finally, the skills associated with Comprehension, with a total frequency of 47.

In order to better explain the numbers, consider the example of the skill "create". It can be found in the Information Usage category and has a frequency of 16. This means that "create" is mentioned in 16 of the selected articles. As shown in Figure 3, the skills found in the Evaluation, Information Usage, and Identification groups have similar frequencies, whereas the skills under the Comprehension group are less frequent in the selected literature. This is an interesting fact, taking into account that this group of skills is key to achieving high IL levels or results.

**RQ3. What tools and techniques are proposed in the literature for assessing information skills of school students?**

Figure 4 summarises the most frequent assessment tools. 31 studies used custom made questionnaires to assess information skills on students, tailored to their demographic target, since such instruments allow the easy gathering of quantitative data and can be simultaneously applied to a large group of people.

Another frequently used approach were individual interviews (13 studies). Although, this method allows the gathering of rich data for a qualitative study, it lacks the massiveness of questionnaires.

Some studies apply both questionnaires and interviews (Herring, 2006; Adhikari, Mathrani & Scogings, 2016; Kovalik, Yutzey & Piazza, 2013; Elster, 2010; Søvik, 2014; Lorenz, Endberg & Bos, 2019; Crary, 2019; Chu, Tse & Chow, 2011) to assess information skills on groups of subjects.

Another relevant tool used for this task is project evaluation (Chung & Kim, 2007; Yu et al., 2011; Pietikäinen, Kortelainen & Siklander, 2017; van Dijk & Lazonder, 2016; Chu, Tse, & Chow, 2011), where students had to develop a research project using both digital and library resources to solve a complex problem, and the entire process was then assessed by teachers and/or researchers. In this kind of evaluation, rich data from the process of information gathering can be obtained, often considering a longitudinal research design.

Case studies were also identified as a relevant tool (Martínez-Piñeiro, Gewerc & Rodríguez-Groba, 2019; Bowler, Large & Rejskind, 2001), along with classroom observation (Adhikari, Mathrani & Scogings, 2016), public discussion forums (Bebbington & Vellino, 2015) and direct participant observation (Aillerie, 2019). Not so frequently, studies also employed standardised tests such as TRAILS (Baji et al., 2018), JMU Information Literacy Test (Smith et al., 2013), ICILS (Zhu et al., 2019), and ICAI (Stanoevska-Slabeva et al., 2015).

Also, two of the literature reviews found (i.e., Hollis (2018) and Mahmood (2017)) cover the evaluation of information skills assessment tools.

The widespread use of custom questionnaires shows the lack of standardisation on this kind of instruments for the K-12 segment, thus making it necessary to build a personalised question set for each study, considering factors such as language, demographic characteristics, cultural aspects, and school level of the target students. Although efforts have been made to make standardised tests for multiple demographics, like the TRAILS (Schloman & Gedeon, 2007) and ICILS (Fraillon & Ainley, 2011) tests, only one study for each of these tests passed the systematic filtering used on this study. This may indicate an opportunity to increase research in the elaboration and application of standarized IL assessment tests for K-12.

**RQ4. What technological tools are discussed in the literature to develop information skills on school students?**

Most of the selected articles did not focus on accurately reporting the technological tools used in their studies. Some indicated in general terms using search tools and ICTs, such as library databases and web browsers (Stanoevska-Slabeva et al., 2015; Siddiq et al., 2016; Azura-Mokhtar, Majid & Foo, 2007; Aillerie, 2019).

Three of the selected articles specifically report the usage of technological tools to develop information skills in students:

- Claro et al. (2012, p.1045) reported that "...a piece of software was developed that simulated ICT applications and the tasks designed emulated real-life school work situations". The tool consisted of an environment with commonly used applications: a virtual desk, an email administrator, an Internet browser, a text processor, a spreadsheet, and a program to design presentations. Additionally, there was a chat window where a virtual conversation among three young classmates was simulated, during which they were asked to perform different tasks.
- Bebbington and Vellino (2015) uses the Minecraft game to teach Information Literacy skills to teenagers. The authors report that Minecraft encourages

- players to assess the information available to make decisions, recognise the need for information, identify the sources that will provide the information needed, find and evaluate information (either alone or in collaboration), apply it, and finally evaluate the results.
- van Dijk and Lazonder (2016, p.196) describe the use of the SCY-Lab learning environment, which "enables students to learn about science topics by solving authentic problems through inquiry and collaboration". With this tool, students take on the role of a scientist working on applied research. Learning is generated from the constructionist paradigm through the creation, discussion, and reuse of learning objects generated by other students.

**RQ5. How does the use of technological tools impact on the development of information skills?**

The answer to this question is closely related to the previous one. Since not many technological tools were reported, the information regarding their impact on the development of online search skills is scarce. However, there are a couple of findings that are worth mentioning:

- Bebbington and Vellino (2015) comment that the game effectively led players to search for information, evaluate the information shared by other players or looking in other spaces but maintaining attention to the source and the relevance about their needs in the game.
- van Dijk and Lazonder (2016) do not discuss the impact of using the tool. However, it is suggested for future works to focus the research on learning objects created by the students within their science project. Other authors (Aillerie, 2019; Stanoevska-Slabeva et al., 2015) comment on the positive impact of the students' usage of technological tools on the experience and motivation in teaching and learning practices.

**RQ6. Which IL standards for k-12 students are addressed in the literature?**

The selected articles mention 26 IL oriented sets of standards, models, and frameworks. The full list is presented in Table 4, where items were sorted by their frequency in the selected articles. The Scope column indicates whereas the item addresses Information Literacy (IL), Digital Competences (DC) or the Search Process (S). The Education Level column shows the target population considered when designing the item, and the Region column indicates its country of origin. The Cited In column indicates the articles that mention each of the tools.

18 of the standards, frameworks and models listed in Table 4 specifically address IL. They can be classified according to their target population:

- Elementary school students: Super 3 (Eisenberg & Berkowitz, 1990) and ILPO (Ryan & Capra, 2001).
- Secondary school students: PLUS model (Ryan & Capra, 2001).
- K-12 students: ALA/AASL Standards for the 21st-Century Learner (American Association of School Librarians, 2009), IFLA/UNESCO standards (Sætre & Willars, 2002), ALA/AECT Information Power framework (American Association of School Librarians and Association for Educational Communications and Technology, 1998), 6+3 model (Mokhtar et al., 2009), i-Competent model (Foo, Majid & Chang, 2017), Information Literacy Standard for Primary and Secondary Students (Zhu et al., 2019), Irving's information skills model (Irving, 1985), New South Wales' information process (School Libraries, Learning systems, State of New South Wales (Department of Education), 2015) and Focus on Inquiry model (Oberg, 2004).

As detailed in Table 4, other standards, frameworks and models are mentioned in the selected articles. Most of them are oriented towards higher education or a general audience, and some focus on information search and information inquiry in a general sense. However, Super 3 is a variant for younger students of The Big Six (Eisenberg & Berkowitz, 1990), which considers secondary and higher education. In the context of digital competences, (Norwegian Directorate for Education and Training, 2013) considers secondary education.

It is interesting to note that, of the K-12 oriented standards, only the ALA/AASL Standards for the 21st-Century Learner (American Association of School Librarians, 2009) were mentioned 5 or more times. This suggests that most research and efforts have been oriented towards higher education students, or, at least, there is more consensus of what IL competences this group of students should develop. It is also interesting that English-speaking countries lead the creation of IL frameworks and models, specially USA, Australia, New Zealand, and UK, and that these are perceived as the most relevant in the selected literature.

Another remarkable fact is that most of the selected articles mention standards, models and frameworks designed for higher education or for a general audience, revealing that further work is needed to develop and promote such tools for K-12 students.

**Discussion**

As we have shown through this article, research on Information Literacy is intensive and broad given its relevance in education contexts. The number of studies in this area grows every year, which is aligned with the development of digital resources.

There is great consensus related to the concept of information literacy and the skills that users need to develop in order to be considered literate. Although the essential skills related to the process of receiving and producing information, remain the same, the concept and skills related to IL have evolved over the years as technology improves and spreads out. Thus, new aspects that have emerged as demands of the evolution of ICTs have been incorporated, such as critically analysing the information, being creative, inventive, solving problems as well as being able to explore the potential of the Internet and digital devices.

Institutions such as the American Library Association (ALA) and the Association of College & Research Libraries (ACRL) point out the need to develop IL skills on individuals who have to face this modern world's environment full of diverse information choices: "The sheer abundance of information will not in itself create a more informed citizenry without a complementary cluster of abilities necessary to use information effectively" (American Library Association and American Library Association for College and Research Libraries, 2000, p.2). There are many and complex skills to be developed, but there are still not many applications, technological tools, nor a wide variety of proposals that could be used to help the population to do so.

Models such as The Big6, have shown to have a positive impact on students' IL skills levels and improve their comprehension and problem-solving skills (Baji et al., 2018; Chen, Chen & Ma, 2017). This particular model helps students to develop Information Literacy skills following some steps to solve problems. Other models, such as Kuhlthau's ISP, also present some positive impacts on the students' development of IL skills (Herring, 2006; Chang et al., 2012; Pietikäinen, Kortelainen & Siklander, 2017; Shannon, Reilly & Bates 2019; Aillerie, 2019; Chu, Tse & Chow, 2011).

Although it is true that environments to evaluate information skills, such as ROSS for university students (Partridge et al., 2008), NEURONE for school students (González-Ibáñez et al., 2017), or the tool used in PISA (OECD, 2015), as well as tools to develop these skills, such as Google Lesson Plans and Google a Day (Leu et al., 2013) exist, we were unable to find, among the studies selected for this review, software or applications that would help people develop those skills without the mediation of a teacher or library staff.

Our article selection focuses primarily on K-12 students, and, to a lower extent, on teachers and other educators such as librarians. Nevertheless, research has also been conducted on information literacy related to other age groups such as children, adults, undergraduate students, graduate students and the elderly.

The most commonly used methods for data collection are questionnaires and interviews. In addition to them, the evaluation of projects proposed to students is also applied. Although less frequently used, project evaluation has shown to bring fruitful results. Case studies, detailed observation of task accomplishment, and the examination of learning processes and other skills related to Information Literacy generate important results that may have an impact on helping many people to develop IL competences. For instance, it could lead to the design and implementation of methodologies, applications, and techniques to support this particular learning process.

Even though questionnaires are the most used method to assess IL competences, selected studies do not mention any standard instrument available that could be widely used. A possible explanation for this are cultural differences across countries, among other factors that lead to the adaptation of the original resources or to the development of new ones. Standard questionnaires for IL assessment would make possible to quantitatively compare different studies in this area regardless of the origin country.

Regarding the technological tools used in the studies, we could notice that they are not clearly and sufficiently described. Therefore, very few of the articles selected for this review refer to the impact of using such tools. Claro et al. (2012) used software that simulated

ICT applications, Bebbington and Vellino (2015) used the Minecraft game, and van Dijk and Lazonder (2016) used the SCY-Lab learning environment. All three articles mention the positive effects of these tools in the learning process. This leads to the thought that there is a need to develop and test more technological tools, alongside better descriptions of the tools themselves and their impact.

Selected literature mention and use several standards over youth education. Most of the studies are oriented to high school students and only a few models and standards are oriented to elementary education. Besides that, few models and standards for the population in general are mentioned and proposed. We know that what is expected from young people is usually the same expected from citizens of a community; however, there are many adults and professionals from different fields who need to develop their Information Literacy competences.

The advances in digital information and communication technology raises the necessity to prepare individuals so that they efficiently and responsibly use information and the affordances available, and to help them to learn how to learn (Isfandyari-Moghaddam & Kashi-Nahanji, 2011; American Library Association & American Library Association for College and Research Libraries, 2000). CAUL/ANZIIL, mentioned by Bundy (2004), reminds us that "information literacy initiates, sustains, and extends lifelong learning through abilities that may use technologies but are ultimately independent of them" (p. 4).

**Conclusions**

This systematic review of the literature aims to gather information on the evolution and current state of the development and assessment of IL at school level (K-12). For this study, a set of research questions were asked. Research questions analyse various aspects of the selected papers, namely (1) definitions for IL, (2) skills and competences related to IL, (3) tools and techniques applied in IL assessment at K-12 level, (4) technological tools used to develop IL skills, (5) their impact on students' skills, and (6) the main IL standards, models, and frameworks addressed in the selected literature.

Most of the selected studies focus on secondary school students, indicating a need for further research on this topic over elementary students. Although the majority of the studies included in our review follows a quantitative approach, an important number of studies using mixed methods was found. This indicates that IL development and assessment is a complex phenomenon that cannot be easily assessed excluding individual aspects. It is interesting to note that selected articles not only consider K-12 students, but also teachers and librarians. Another remarkable finding is that the amount of research regarding IL at school level has continued to increase over the last years. This may be linked to the increasing number of people able to access information via digital technologies.

From the selected studies for this review, we acknowledge that IL has been well defined (RQ1) as "the ability to locate, access, search, evaluate and use information in different contexts" (Chang, Foo & Majid, 2014, p.1), whether it is in physical or digital environments. This definition is usually taken from well-established standards, but the authors themselves provide similar ones.

We classify the main skills related to IL (RQ2) into four major categories: Identification, Evaluation, Information Usage and Comprehension. These categories are aligned with studies such as (Leu et al., 2014), where IL skills are not just used as part of the learning process, but also integrate synthesising and communicating the acquired knowledge. It is important to notice that the skills developed by K-12 students are not well balanced, since some of them are more developed than others. Comprehension skills such as Understand, Process, Organise and Synthesise, among others, for instance, deserve more attention from educators in order to be developed.

Most of the tools and techniques used to assess IL in the selected studies (RQ3) are questionnaires, interviews and project evaluations. Most of them use customised tools according to the region and demographic characteristics of the students, showing a lack of standardisation in the assessment process.

Selected studies did not make an extensive report about the technological tools used to develop IL skills on students (RQ4). Only three studies made a detailed report of their tools and experiences (Bebbington & Vellino, 2015; Claro et al., 2012; van Dijk and Lazonder, 2016), although their impact on the development of IL (RQ5) is not clearly stated.

As the analysed studies indicate, the most productive and frequently used methodology for IL analysis is a mixed one, which deals with both quantitative and qualitative data. Most questionnaires are customised, thus evidencing the absence of a standardised questionnaire that meets the demands of research carried out in this area. Such an assessment tool could be widely adopted. We understand that it is very difficult to develop such an instrument because of idiomatic and cultural differences. Nevertheless, we were unable to find an attempt to, at least, provide orientations on how to construct a suitable questionnaire.

The studies analysed in this systematic review expose the lack of tools to promote the development of IL skills. This is an unique opportunity for information professionals and researchers, who could propose specialised tools to support students in this development process. It must be noted that this type of tools exists (e.g., ROSS (Partridge et al., 2008), NEURONE (González-Ibáñez et al., 2017), and Buscador Escolar (Ibieta, Hinostroza, and Labbé 2019)), however, the articles in which they are introduced did not pass the selection criteria. Certainly, the development of such tools is not enough, but its evaluation is fundamental to determine their impact on the development of information skills.

  Although this literature review was conducted rigorously, there are some limitations to this study. First, relevant articles may have been excluded, since (1) the search was limited only to English and Spanish languages, (2) not all article databases were consulted because of access restrictions and (3) there were access limitations for some of the articles that passed the abstract filter. Second, the lack of consensus for the term Information Literacy in Spanish could have limited the incorporation of all possible meanings in the query string.

  We expect this study to contribute to the understanding of the impact of IL on K-12 students, the importance of IL skills and the need to develop them as early as possible, since students will live and work in a world where information is ever-increasing, but not necessarily accurate or trust-worthy. Nevertheless, IL can be seen as the tip of an iceberg, since it is also related to computer skills, reading comprehension and problem solving. Therefore, IL skills should be taught integrated into the main academic curricula, and also incorporating the use of computers or technological devices in general.

  Among the results of this review we found evidence of the need for new assessment instruments and learning tools. The former, which comprises both an agreed set of skills and properly defined means to evaluate them, will allow the conduction of comparable studies despite the differences due to cultural or regional contexts. The latter, to be used in all different educational levels, also needs to be developed, aiming to improve the effectiveness of the learning process on students, since they could take advantage of the educational resources available.

Information literacy development and assessment at school level    31

**Footnotes**

[1] Original in Spanish. Free translation by the authors.

[2] Original in Spanish. Free translation by the authors.

**Tables**

Table 1

Considered works after each stage of the selection process. Accepted (discarded) in cells.

| Source | Retrieved | Remove duplicates | Filter by title | Filter by abstract | Downloaded | Filter by quality |
|---|---|---|---|---|---|---|
| ECIL | 37 | 20 (17) | 7 (13) | 3 (4) | 3 (0) | 0 (3) |
| ICIL | 3 | 3 (0) | 2 (1) | 0 (2) | 0 (0) | 0 (0) |
| JIL | 107 | 104 (3) | 54 (50) | 13 (41) | 13 (0) | 5 (8) |
| Scielo | 38 | 18 (20) | 7 (11) | 4 (3) | 4 (0) | 0 (4) |
| Scopus | 1444 | 604 (840) | 171 (433) | 117 (54) | 81 (36) | 19 (62) |
| WOS | 749 | 485 (264) | 199 (286) | 115 (84) | 108 (7) | 29 (79) |
| **Total** | **2378** | **1234 (1144)** | **440 (794)** | **252 (188)** | **209 (43)** | **53 (156)** |

Table 2

Inter-rater reliability for each quality assessment criterion.

| QA | Agree percentage | Kappa | Statistic |
|---|---|---|---|
| 1 | 84.135 | 0.642 | 9.298 |
| 2 | 82.692 | 0.643 | 9.276 |
| 3 | 87.981 | 0.750 | 10.866 |
| 4 | 83.173 | 0.650 | 9.387 |
| 5 | 84.615 | 0.612 | 9.990 |
| 6 | 87.019 | 0.738 | 10.668 |

Table 3

Selected articles ordered by author's lastname

| Author(s) | Year | Author(s) | Year |
|---|---|---|---|
| Adhikari et al. | 2016 | Isfandyari-Moghaddam and Kashi-Nahanji | 2011 |
| Aillerie | 2019 | Kankam and Nsibirwa | 2019 |
| Akkoyunlu and Yilmaz | 2011 | Kovalik et al. | 2013 |
| Álvarez and Merce | 2015 | List et al. | 2020 |
| Azura-Mokhtar et al. | 2007 | Lorenz et al. | 2019 |
| Baji et al. | 2018 | Mahmood | 2017 |
| Bebbington and Vellino | 2015 | Majid et al. | 2020 |
| Bielba-Calvo et al. | 2015 | Martínez-Piñeiro et al. | 2019 |
| Bielba-Calvo et al. | 2017 | Ngo et al. | 2019 |
| Bowler et al. | 2001 | Nunaki et al. | 2019 |
| Chang et al. | 2014 | Pietikäinen et al. | 2017 |
| Chen and Chen | 2016 | Scherer and Siddiq | 2015 |
| Chen et al. | 2017 | Shannon et al. | 2019 |
| Chu et al. | 2011 | Siddiq and Scherer | 2016 |
| Chung and Kim | 2007 | Siddiq et al. | 2016 |
| Claro et al. | 2012 | Smith et al. | 2013 |
| Bebbington and Vellino | 2015 | Majid et al. | 2020 |
| Crary | 2019 | Søvik | 2014 |
| Dotan and Aharony | 2008 | Stanoevska-Slabeva et al. | 2015 |
| Elster | 2010 | Stanoevska-Slabeva et al. | 2017 |
| Foo et al. | 2017 | Tachie-Donkor and S. | 2017 |
| Gerick et al. | 2017 | van Dijk and Lazonder | 2016 |

| Author(s) | Year | Author(s) | Year |
|---|---|---|---|
| Grimley | 2012 | Yu et al. | 2011 |
| Guthrie and Klauda | 2014 | Chang et al. | 2012 |
| Haras et al. | 2008 | Zhang and Zhu | 2016 |
| Heinström | 2006 | Zhou and Lam | 2019 |
| Herring | 2006 | Zhu et al. | 2019 |
| Hollis | 2018 | | |

Table 4

Main IL-related standards, models and frameworks.

| Standard, framework or model | Scope | Ed. Level | Region | Mentions | Cited in |
|---|---|---|---|---|---|
| The Big 6 | IL | General | USA | 8 | Herring (2006); Chang et al. (2014); Foo et al. (2017); Chang et al. (2012); Chen and Chen (2016); Baji et al. (2018); Chen et al. (2017); Pietikäinen et al. (2017) |
| Kuhlthau's ISP model | SP | General | | 6 | Herring (2006); Chang et al. (2012); Pietikäinen et al. (2017); Shannon et al. (2019); Aillerie (2019); Chu et al. (2011) |
| ACRL/ALA Information Literacy for Higher Education | IL | Higher | USA | 5 | Majid et al. (2020); Bielba-Calvo et al. (2015); Smith et al. (2013); Zhu et al. (2019); Bielba-Calvo et al. (2017) |
| CAUL/ANZIIL | IL | Higher | Australia & New Zeland | 5 | Majid et al. (2020); Bielba-Calvo et al. (2015); Yu et al. (2011); Zhu et al. (2019); Bielba-Calvo et al. (2017) |
| SCONUL 7 pillars | IL | Higher | UK | 5 | Majid et al. (2020); Chang et al. (2014); Bielba-Calvo et al. (2015); Zhu et al. (2019); Bielba-Calvo et al. (2017) |
| ALA/AASL Standardsfor the 21st-Century Learner | IL | K-12 | USA | 5 | Chen and Chen (2016); Bielba-Calvo et al. (2015); Ngo et al. (2019); Dotan and Aharony (2008); Bielba-Calvo et al. (2017) |
| ISTE National Educational Technology Standards | DC & IL | General | International | 4 | Bielba-Calvo et al. (2015); Álvarez and Mercè (2015); Bielba-Calvo et al. (2017); Siddiq et al. (2016) |
| IFLA/UNESCO | IL | K-12 | International | 4 | Álvarez and Mercè (2015); Yu et al. (2011); Zhu et al. (2019); Elster (2010) |
| DigComp | DC & IL | General | Europe | 4 | List et al. (2020); Martínez-Piñeiro et al. (2019); Lorenz et al. (2019); Siddiq et al. (2016) |

| Standard, framework or model | Scope | Ed. Level | Region | Mentions | Cited in |
|---|---|---|---|---|---|
| ALA/AECT Information Power | IL | K-12 | USA | 3 | Herring (2006); Majid et al. (2020); Bowler et al. (2001) |
| ICT strategic framework for education | DC & IL | General | New Zeland | 2 | Adhikari et al. (2016); Bielba-Calvo et al. (2017) |
| CRUE-TIC/REBIUN | DC & IL | Higher | Spain | 2 | Bielba-Calvo et al. (2015, 2017) |
| 6+3 | IL | K-12 | Singapore | 2 | Foo et al. (2017); Chang et al. (2012) |
| i-Competent | IL | K-12 | Singapore | 2 | Majid et al. (2020); Foo et al. (2017) |
| UDIR Norwegian Directorate for Education and Training | DC & IL | K-12 | Norway | 2 | Søvik (2014); Siddiq et al. (2016) |
| Super 3 | IL | Primary | USA | 2 | Foo et al. (2017); Chen and Chen (2016) |
| Stripling's Inquiry Model | SP | General |  | 2 | Herring (2006); Chang et al. (2012) |
| 7 Faces of Information Literacy | IL | Higher | Australia | 1 | Chang et al. (2014) |
| Six Frames for Informed Learning | IL | Higher | Australia | 1 | Chen and Chen (2016) |
| Information Literacy Standard for Primary and Secondary Students | IL | K-12 | China | 1 | Zhu et al. (2019) |
| Irving's information skills model | IL | K-12 | UK | 1 | Chang et al. (2012) |
| New South Wales' information | IL | K-12 | Australia | 1 | Chang et al. (2012) |
| Focus on Inquiry | IL | K-12 | Canada | 1 | Herring (2006) |
| ILPO | IL | Primary | Australia | 1 | Herring (2006) |
| PLUS | IL | Secondary | UK | 1 | Herring (2006) |
| Wilson's model of information behavior | DC & IL | General |  | 1 | Kankam and Nsibirwa (2019) |

Figures

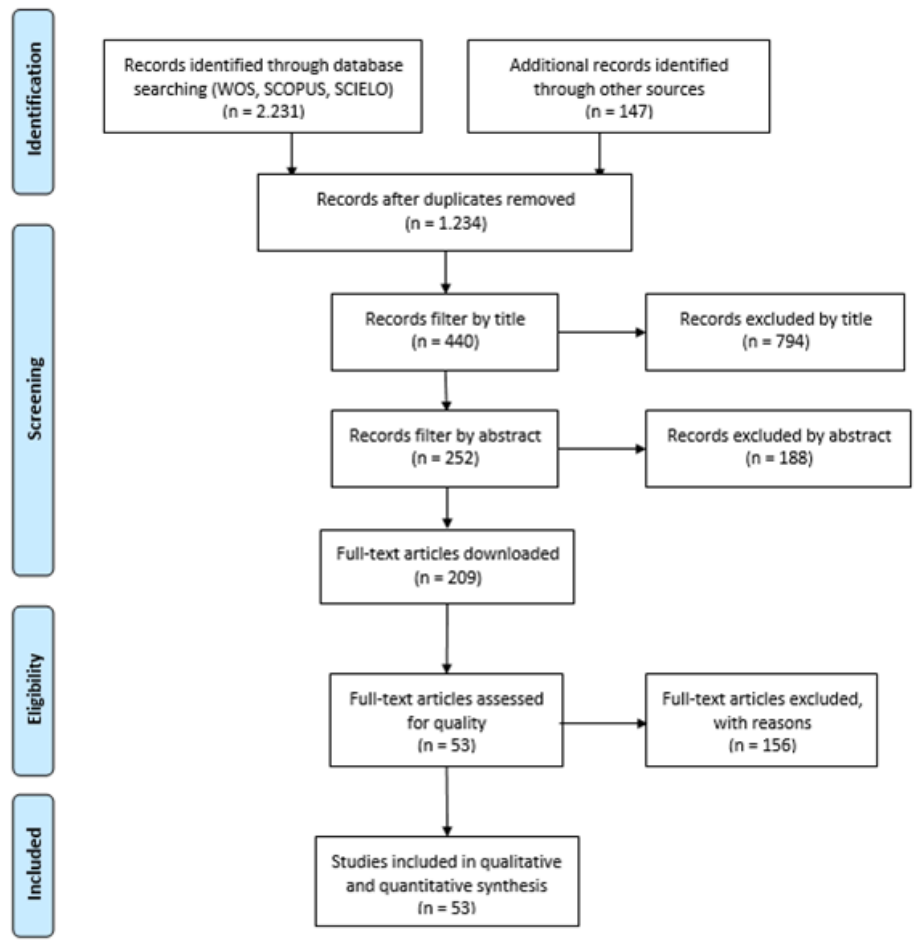

*Figure 1*. PRISMA flow diagram for the study selection process (Page et al. (2021)).

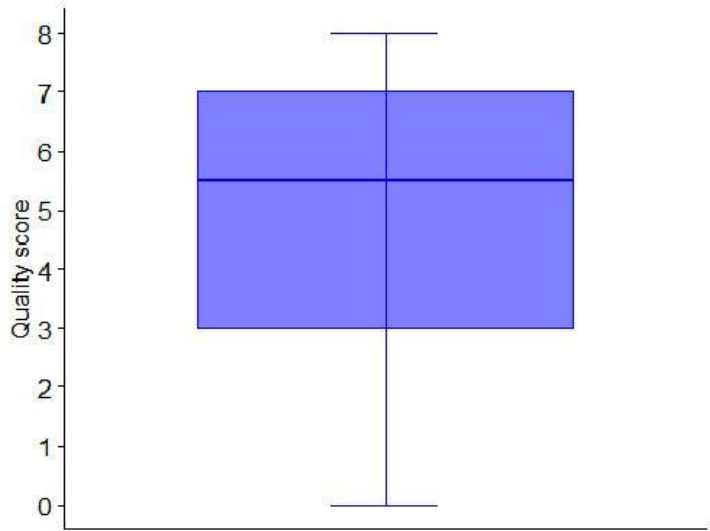

*Figure 2*. Quality scores distribution of selected papers.

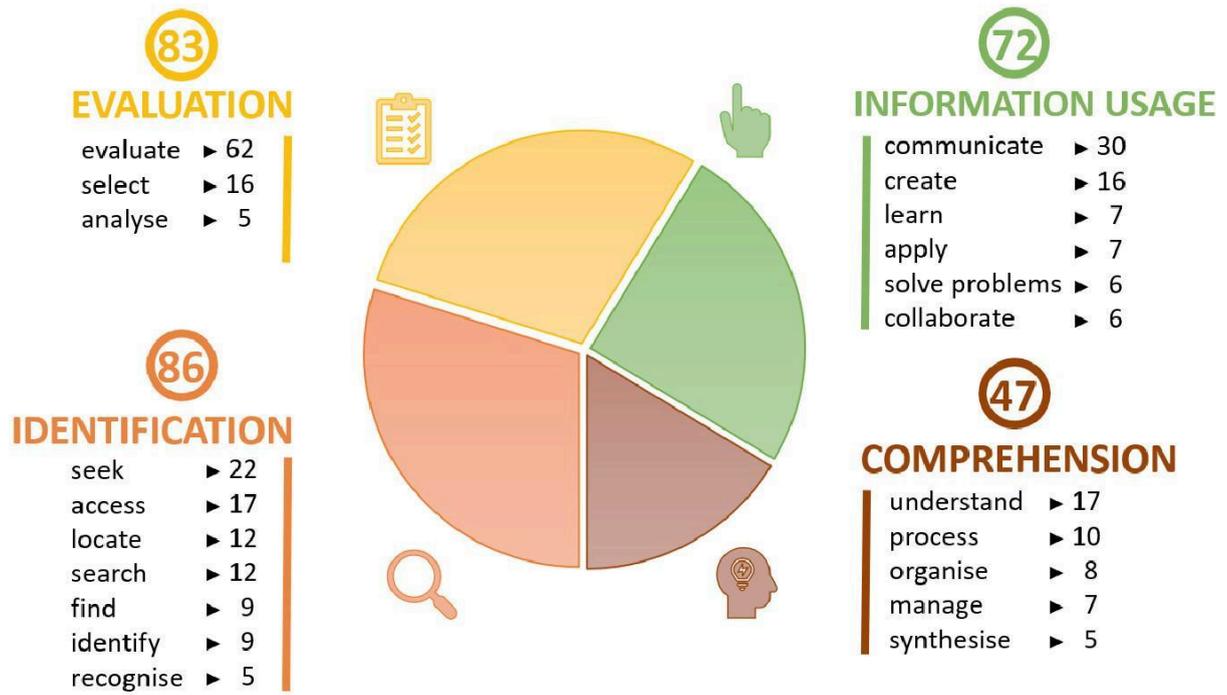

*Figure 3*. Skills frequencies, grouped by similarity.

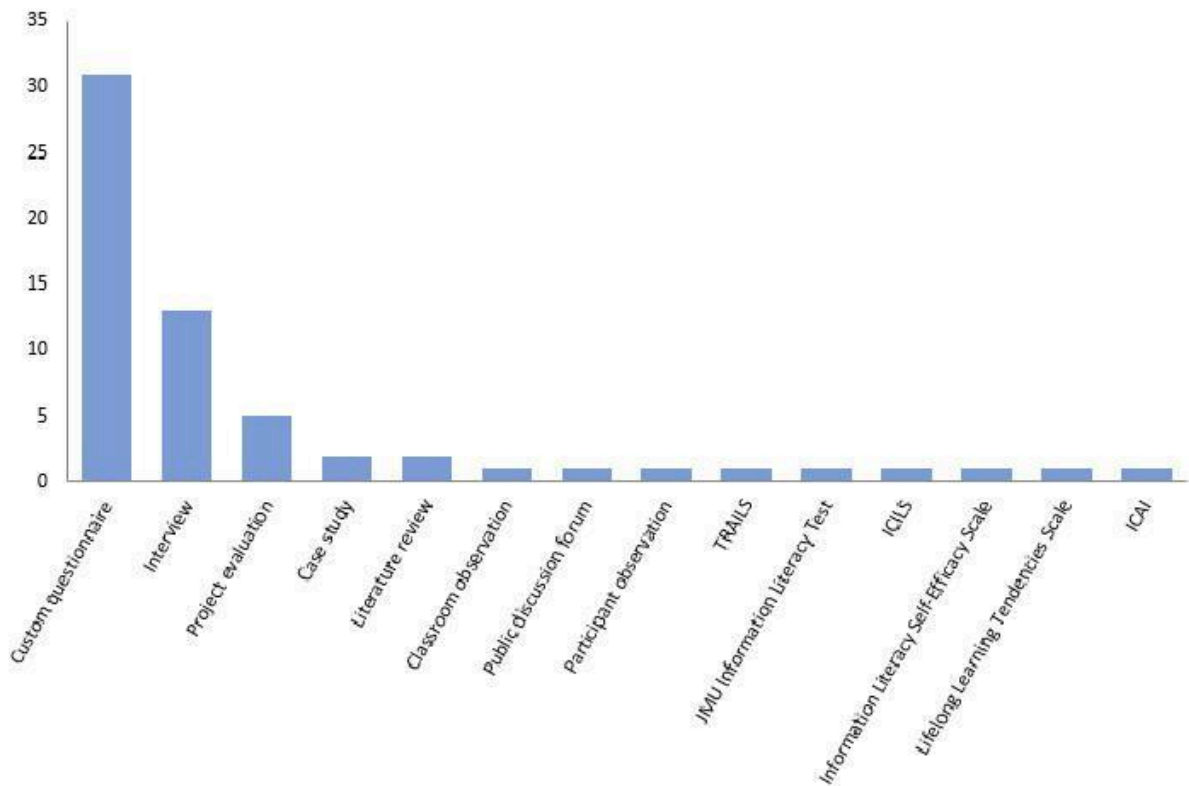

*Figure 4*. Frequencies of appearance of tools and techniques for assessing information skills of school students.